\documentclass[preprint,12pt]{elsarticle}

\usepackage{graphicx}
\usepackage{amsmath,amssymb,amsfonts}
\usepackage{url}
\usepackage{textcomp}
\usepackage{float}
\usepackage{booktabs}
\usepackage{xcolor}
\usepackage{setspace}
\usepackage{colortbl}
\usepackage{multirow}
\usepackage{tikz}
\usepackage{array}

\usepackage{newtxtext}
\usepackage{newtxmath}

\usepackage{tikz}
\usetikzlibrary{positioning,arrows,trees}
\usepackage[edges]{forest}

\begin{document}

\onehalfspacing  

\begin{frontmatter}

\title{A Hybrid Machine Learning Approach for Graduate Admission Prediction and Combined University--Program Recommendation}

\author{Melina Heidari Fard$^{1}$}
\ead{melinaheydarifard@gmail.com}
\author{Elham Tabrizi$^{2}$\corref{cor1}}
\ead{elham.tabrizi@khu.ac.ir}

\cortext[cor1]{Corresponding author}

\address{Department of Mathematics, Faculty of Mathematical Sciences and Computer, Kharazmi University, Tehran, Iran}

\begin{abstract}
Graduate admissions have become increasingly competitive. This study highlights the need for a hybrid machine learning framework for graduate admission prediction, focusing on high-quality similar applicants and a recommendation system. The dataset, collected and enriched by the authors, includes 13,000 self-reported GradCafe application records from 2021 to 2025, enriched with features from the OpenAlex API, QS World University Rankings by Subject, and Wikidata SPARQL queries. A hybrid model was developed by combining XGBoost with a residual refinement k-nearest neighbors module, achieving 87\% accuracy on the test set. A recommendation module, then built on the model for rejected applicants, provided targeted university and program alternatives, resulting in actionable guidance and improving expected acceptance probability by 70\%. The results indicate that university quality metrics strongly influence admission decisions in competitive applicant pools. The features used in the study include applicant quality metrics, university quality metrics, program-level metrics, and interaction features.

\end{abstract}


\end{frontmatter}

\section{Introduction and Related Work}
\label{S:1}

Graduate admissions have become increasingly competitive due to the global expansion of postgraduate education and the growing number of highly qualified applicants competing for a limited number of positions. In many selective programs, applicants exhibit nearly indistinguishable academic profiles, with comparable undergraduate GPAs, standardized test scores, coursework rigor, and research or internship experience. This high degree of similarity makes accurate admission prediction challenging and limits the usefulness of many existing decision support tools for applicants near the acceptance boundary \cite{Huang2015,Yoo2017}.

To address this challenge, prior studies have formulated graduate admission prediction as a supervised learning problem using structured applicant features \cite{Huang2015,Yoo2017}. Subsequent research has demonstrated that nonlinear models such as random forests, support vector machines, and gradient boosting methods provide superior predictive performance \cite{Kumar2019,Breiman2001,Cortes1995}. In particular, gradient boosted decision trees, including XGBoost and LightGBM, have emerged as strong baselines \cite{Chen2016,Ke2017}. Academic performance indicators consistently appear as the most influential predictors \cite{Jayaprakash2014,Kuncel2001}. However, many existing studies rely on small or institution specific datasets, limiting their generalizability \cite{Delen2011}. Larger self reported datasets such as GradCafe partially address this limitation, yet predictive performance remains limited for applicants with highly similar profiles \cite{GradCafe,OpenAlex,QS,Wikidata}.

Interpretability has become an important consideration in educational decision support systems, motivating the adoption of methods such as SHAP and LIME \cite{Lundberg2017,Ribeiro2016,Molnar2022}. Furthermore, hybrid recommendation systems, which combine multiple approaches to improve guidance for users, have been widely studied \cite{Burke2002,Ricci2015}. Nevertheless, careful consideration of data missing not at random is crucial when training and evaluating these systems \cite{Steck2010}. Most existing approaches focus primarily on binary admission prediction and offer limited guidance for applicants who are likely to be rejected. In particular, applicants near the decision boundary remain underexplored, despite being the group most in need of realistic and actionable recommendations.

In this work, we address these limitations by constructing a large, global, multi source dataset of approximately 13,000 competitive graduate admission records enriched with institutional and program level metadata from OpenAlex, QS World University Rankings by Subject, and Wikidata \cite{OpenAlex,QS,Wikidata}. Building on this dataset, we conduct a comprehensive evaluation of multiple machine learning models under competitive boundary conditions and demonstrate the limitations of single model approaches in high similarity applicant populations. We further propose a hybrid residual refinement framework that combines XGBoost with a k-nearest neighbors (kNN) module to specifically target borderline admission cases, achieving a predictive accuracy of 87\%. Finally, we introduce a dual axis recommendation system that provides institution and program level alternatives for rejected applicants, offering realistic and actionable guidance under real world academic constraints \cite{Burke2002,Ricci2015,Steck2010}.

\section{Data and Preprocessing}
\label{S:3}

\subsection{Dataset Collection}
The dataset used in this study was collected from GradCafe, an online platform where applicants voluntarily report the outcomes of their higher education applications. The initial web scraping process covered application records spanning from 2001 to 2025. An exploratory temporal analysis indicated that records from 2021 onward provide the most representative and stable view of contemporary graduate admission dynamics.

\begin{figure}[htbp]
    \centering
    \includegraphics[width=0.8\textwidth]{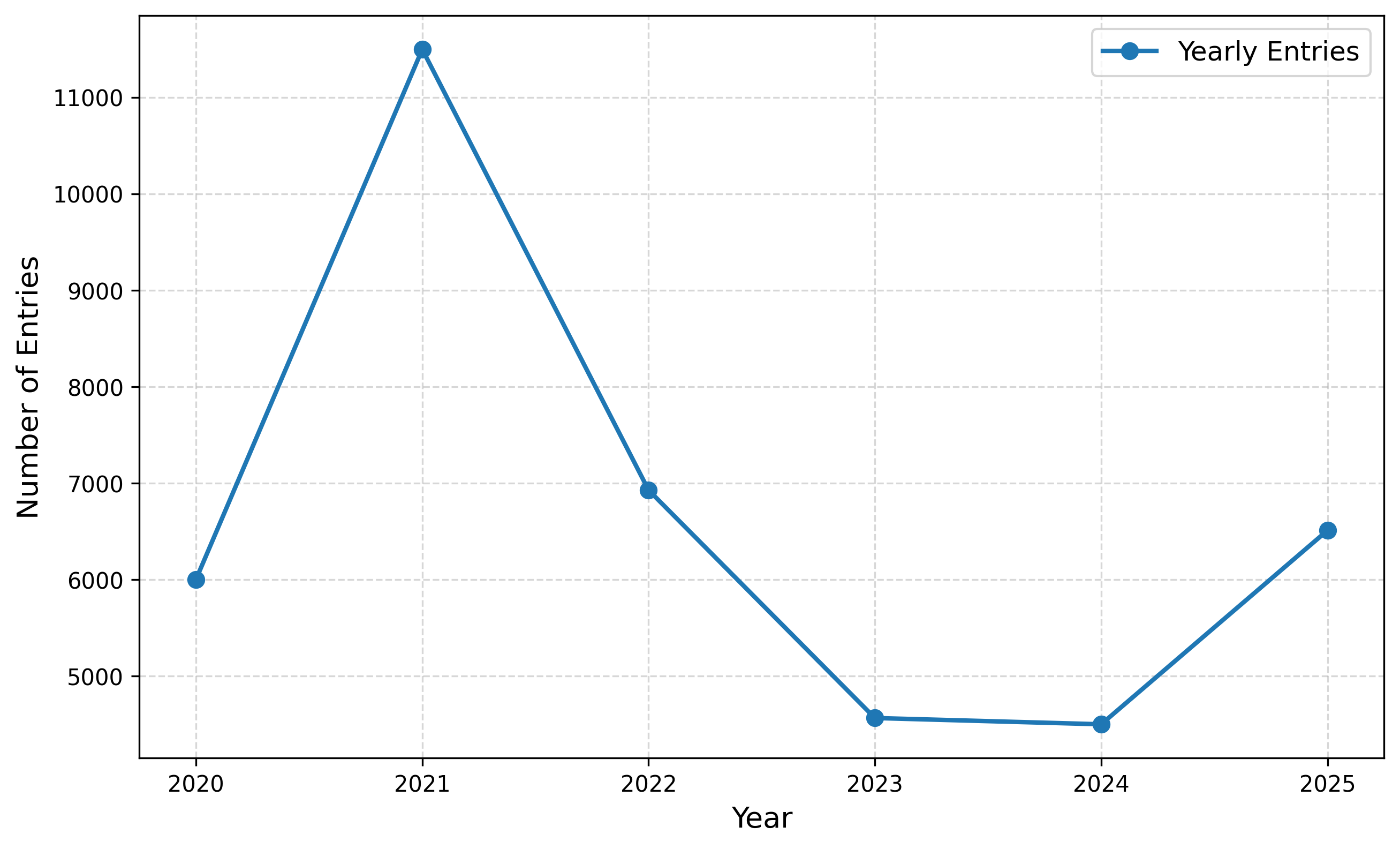}
    \caption{Number of self reported GradCafe application records by year.}
    \label{fig:gradcafe_years}
\end{figure}

In particular, the year 2021 exhibits a pronounced peak with 11{,}497 entries, representing a 91.62\% increase compared to 2020 and accounting for 43.75\% of all cumulative observations. Subsequent years show notable fluctuations, including a decline of 39.76\% in 2022, a partial recovery of 34.12\% in 2023, and a modest rebound in 2025. To balance data volume with temporal relevance, this study restricts the analysis to applications submitted from 2021 onward.

The raw dataset was collected using a custom web crawler implemented in Python with the Scrapy framework. No external APIs or third party data services were used; all information was extracted directly from HTML pages using CSS selectors. A dedicated Scrapy pipeline was employed to perform data validation, cleaning, and transformation into a structured format suitable for downstream analysis. The crawler’s operation depended on the underlying HTML structure of the website, implying that minor code adjustments may be required in response to layout changes. To comply with site usage policies and ensure efficient data collection, the crawler incorporated randomized download delays, custom request headers, a concurrency limit of 32 requests, and an auto throttling mechanism that dynamically adjusted crawling speed based on server response.

The final dataset was stored in JSON Lines (JSONL) format, with timestamped filenames to document the exact date and time of collection. After filtering, the raw dataset contains approximately 34{,}000 records and 14 features, encompassing applicant level attributes such as undergraduate GPA, GRE verbal, analytical writing, and total scores, degree type, and applicant status, application level attributes including target program, university, admission decision, decision date components (day, month, and year), and application season, as well as unstructured user provided notes.

\subsection{University Data Enrichment}
To incorporate institutional context and quality indicators, university-level data were enriched using multiple external sources. After comparing OpenAlex API, Webometrics, and QS World University Rankings, the QS dataset was selected as the primary source due to its completeness and reliability. University names were standardized using normalization procedures and alternative names from OpenAlex, followed by manual verification. This process successfully matched 1,860 universities (98.57\%). The final feature set includes comprehensive QS indicators, covering general information (rank, country, region, focus, research, status), detailed metrics (academic reputation, employer reputation, faculty-student ratio, citations per faculty, international faculty and student ratios, industry income, employment outcomes, sustainability), and the overall score.

\subsection{Program Discipline Mapping}
Program names reported by applicants were heterogeneous and unstandardized. They were manually cleaned and mapped to professional identifiers, with parent disciplines extracted via Wikidata queries. Manual corrections addressed misclassifications and ambiguities, resulting in consistent and interpretable discipline groupings for modeling.

\subsection{Admission Outcome Processing}
The original decision variable included multiple outcomes (accepted, rejected, waitlisted, interview, other, none). Non-final outcomes, representing 27.44\% of observations, were excluded to avoid bias and class imbalance. The remaining data were encoded as a binary variable (accepted = 1, rejected = 0), yielding a nearly balanced distribution of 51\% accepted and 49\% rejected.

\subsection{Feature Selection and Preprocessing}
Correlation analysis among QS indicators revealed redundancy; the most predictive features retained were FSR\_SCORE (faculty-student ratio), CPF\_SCORE (citations per faculty), ISR\_SCORE (international student ratio), and QS\_Rank. Applicant features were processed to handle missingness: GRE scores were excluded due to more than 85\% missingness, while GPA missingness (~50\%) was addressed by dropping approximately 40\% of records and imputing the remaining missing values using the median GPA (3.8), resulting in a total of 2,283 imputed entries. Degree types were filtered to include only PhD and Master’s programs, as all other degrees collectively accounted for only ~4.85\% of observations. Categorical variables (Target\_Program, Target\_Program\_Discipline, Target\_University, Target\_Country) were encoded using frequency encoding to capture relative popularity without introducing high-dimensional sparsity. Numerical features, including Applicant\_GPA, FSR\_SCORE, CPF\_SCORE, ISR\_SCORE, QS\_Rank, and Decision\_Year, were standardized.

\subsection{Feature Engineering Guided by Decision Tree Analysis}
Numerical features including Applicant\_GPA, FSR\_SCORE, CPF\_SCORE, ISR\_SCORE, QS\_Rank, and Decision\_Year were standardized to zero mean and unit variance to ensure comparable contributions across features and improve model performance. During initial baseline experiments, decision tree classifiers revealed a clear hierarchical structure in admission decisions, reflecting real world logic and identifying critical thresholds among applicant and university level features. This insight guided feature engineering and informed the design of interaction features used in subsequent hybrid modeling.

\clearpage  

\begin{figure}[htbp]
\centering
\begin{forest}
for tree={
    draw,
    rounded corners,
    align=center,
    font=\tiny,                
    parent anchor=south,
    child anchor=north,
    edge={->, line width=0.2pt},  
    l sep=8pt,                 
    s sep=2pt,                 
    inner sep=1.5pt,           
    tier/.option=level,
    before typesetting nodes={
        for tree={content/.wrap value={\scalebox{0.9}{##1}}}  
    }
}
[Target\_Degree\_Type $\leq 1.50$,
    edge={->, line width=0.2pt, l=0pt}  
    [Target\_University\_QS\_Rank $\leq -0.39$
        [Target\_Program
            [Target\_University\_QS\_Rank
                [(...)]
            ]
            [Target\_University\_QS\_Rank
                [(...)]
            ]
        ]
        [Target\_Program
            [Target\_University\_FSR\_Score
                [(...)]
            ]
            [Target\_University\_FSR\_Score
                [(...)]
            ]
        ]
    ]
    [Target\_University\_QS\_Rank $> -0.39$
        [Target\_University\_Status
            [Applicant\_Status
                [(...)]
            ]
            [Applicant\_Status
                [(...)]
            ]
        ]
        [Target\_University\_Status
            [Target\_University\_FSR\_Score
                [(...)]
            ]
            [Target\_University\_FSR\_Score
                [(...)]
            ]
        ]
    ]
]
\end{forest}
\caption{Decision tree showing the main splits with numeric thresholds on the first two levels, simplified feature only nodes thereafter, and truncated leaves indicated by `(...)`.}
\end{figure}
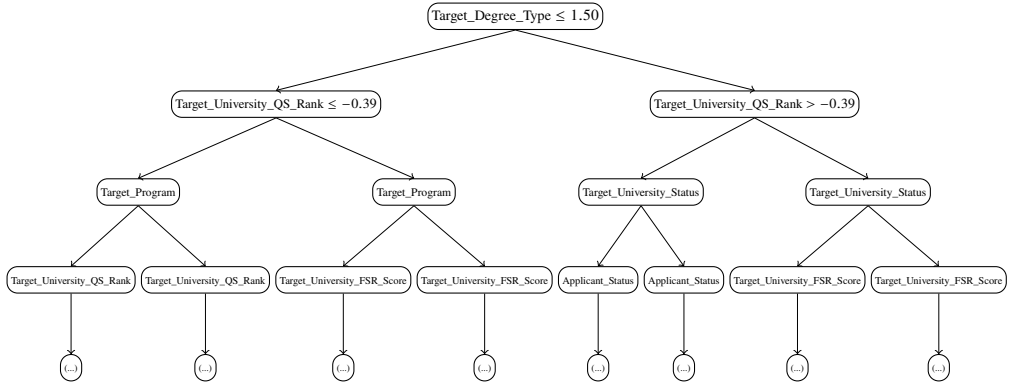

Target\_Degree\_Type emerged as the primary splitter in the decision tree, separating applicants into Master’s ($\leq 1.50$) and PhD ($> 1.50$) regimes.

For Master’s applicants, admission decisions are primarily driven by institutional quality and program level alignment rather than applicant level metrics alone. Target\_University\_QS\_Rank is the earliest and most influential splitter, indicating that university prestige strongly conditions subsequent outcomes. Once institutional quality is established, Target\_Program, Target\_Program\_Discipline, and Applicant\_GPA collectively refine predictions, while sub-dimensions of university reputation, such as FSR, CPF, and ISR, appear repeatedly as secondary refinements. Decision\_Year plays a comparatively minor role.  

For PhD applicants, the decision process emphasizes research oriented and performance sensitive evaluation. University level indicators remain influential, but Applicant\_GPA becomes substantially more decisive. The tree reveals frequent interactions between GPA, university research strength, and program discipline, reflecting a stronger emphasis on academic capability and research fit. Applicant\_Status and Decision\_Year occur more frequently in deeper splits, indicating heightened sensitivity to contextual and temporal factors.  

Overall, university quality metrics consistently emerge as the most influential drivers of admissions outcomes. Applicant\_GPA serves as the second most important signal, while Target\_Program and Target\_Program\_Discipline jointly form the third shared factor. Program\_Admission\_Rate, defined as the proportion of accepted applicants per program, captures real-world program competitiveness, complementing these features. Finally, an interaction term between Program and QS Rank encapsulates how program competitiveness interacts with institutional prestige, providing additional explanatory power beyond individual variables.

\begin{table}[H]
\centering
\arrayrulecolor{blue}
\begin{tabular}{ll}
\hline
\textbf{Feature} & \textbf{Description} \\
\hline
Decision & Accepted / Rejected \\
Applicant\_GPA & Academic performance \\
Decision\_Year & Year the decision was reported \\
Target\_University\_FSR\_Score & Faculty--student ratio \\
Target\_University\_CPF\_Score & Citations per faculty \\
Target\_University\_ISR\_Score & International student ratio \\
Target\_University\_QS\_Rank & Global QS rank \\
Target\_Degree\_Type & Master’s / PhD \\
Applicant\_Status & International / American \\
Target\_University\_Status & Public / Private (for-profit / not-for-profit) \\
Target\_Program & Encoded program \\
Target\_Program\_Discipline & Encoded discipline \\
Target\_University & Encoded university \\
Target\_Country & Encoded country \\
Program\_Admission\_Rate & Acceptance proportion per program \\
Program $\times$ QS Rank & Interaction feature \\
\hline
\end{tabular}
\caption{Description of features used in final model}
\label{tab:feature_description}
\arrayrulecolor{black}
\end{table}

\subsection{Dataset Summary and Bias Considerations}
After preprocessing, the final dataset consists of approximately 13{,}000 applicants described by 15 features. The dataset exhibits a pronounced geographic concentration applications to U.S. universities account for 84.53\% of the observations, while institutions in Canada and the United Kingdom together contribute approximately 9\%. Other countries, including Switzerland, Spain, Italy, Germany, Australia, Korea, and Taiwan, each represent less than 1\% of the total sample. This imbalance suggests that the trained models primarily capture admission patterns prevalent in U.S. institutions, which should be carefully considered when interpreting model predictions and generalizing results to non U.S. contexts.

Applicant GPA values are predominantly concentrated in the mid to high range, reflecting the inherently competitive nature of graduate admissions. This distributional skew indicates that model performance may be less reliable for applicants with comparatively lower GPAs, thereby introducing a potential source of bias. Consequently, prediction outcomes for underrepresented GPA ranges should be interpreted with caution.
\begin{figure}[H]
    \centering
    \includegraphics[width=0.8\linewidth]{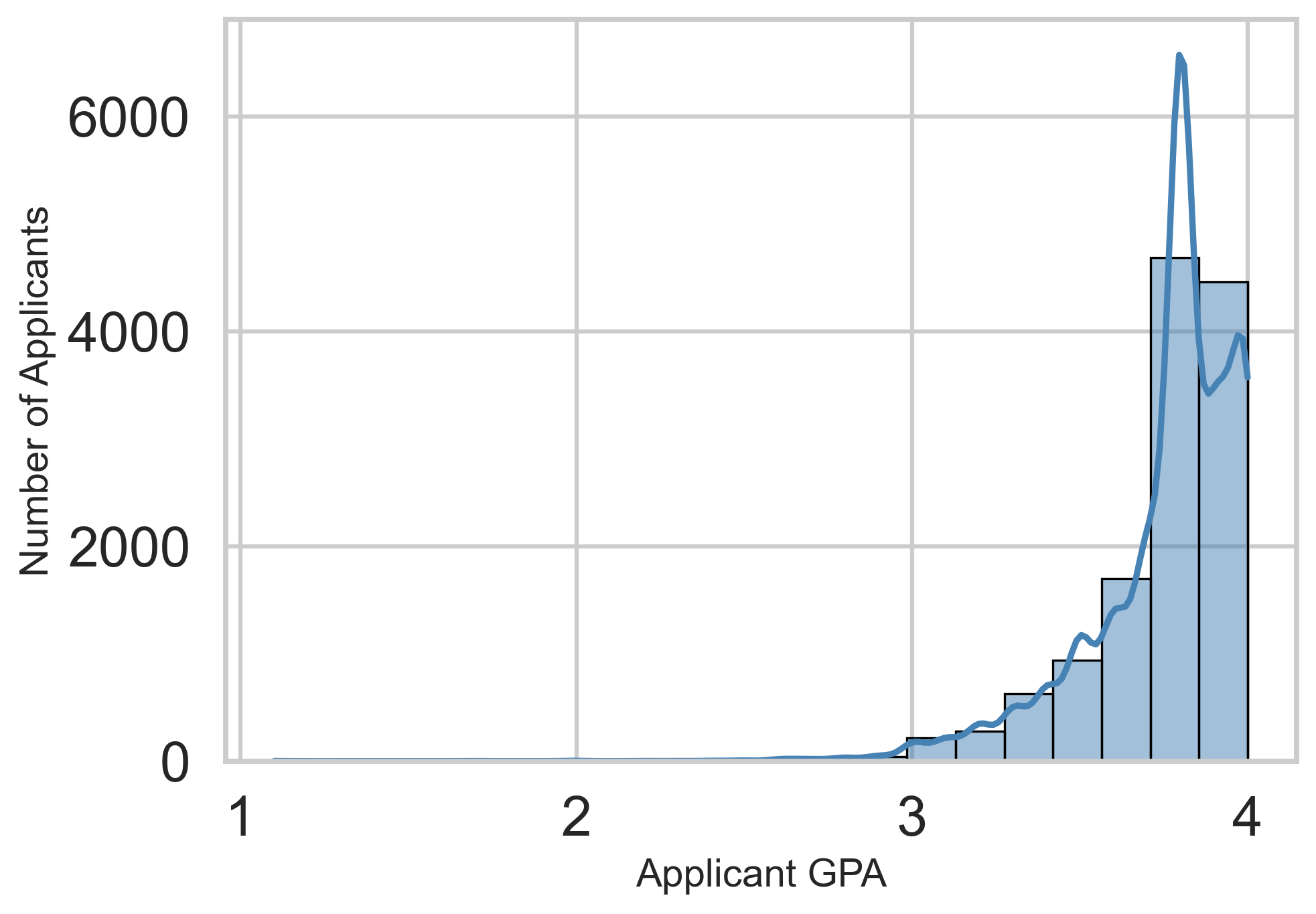}
    \caption{Distribution of applicant GPAs in the final dataset. The histogram shows the number of applicants per GPA bin, and the overlaid line represents the smoothed density.}
    \label{fig:gpa_distribution}
\end{figure}
\begin{table}[H]
\centering
\arrayrulecolor{blue}
\begin{tabular}{lccccccc}
\hline
Feature & Mean & Min & 25th Percentile & Median & 75th Percentile & Max & Skewness \\
\hline
Applicant GPA & 3.752 & 1.10 & 3.68 & 3.80 & 3.90 & 4.00 & $-1.793$ \\
\hline
\end{tabular}
\caption{Summary statistics of applicant GPA.}
\label{tab:gpa_stats}
\end{table}
The applicant dataset shows a diverse distribution across academic programs, with certain fields attracting a higher share of applications. The most represented programs are Computer Science (8.33\%), Food and Resource Economics (8.22\%), and Economics (5.52\%), followed by Physics (4.14\%), Speech-Language Pathology (3.98\%), Chemistry (3.35\%), and Philosophy (3.05\%). Smaller proportions are observed in Public Policy \& Management (2.72\%), English (2.68\%), and Political Science (2.31\%), while the remaining programs collectively account for 55.72\% of the dataset.
Therefore, at the discipline level, certain disciplines attract a higher number of applicants.
\begin{figure}[H]  
    \centering
    \includegraphics[width=0.95\linewidth]{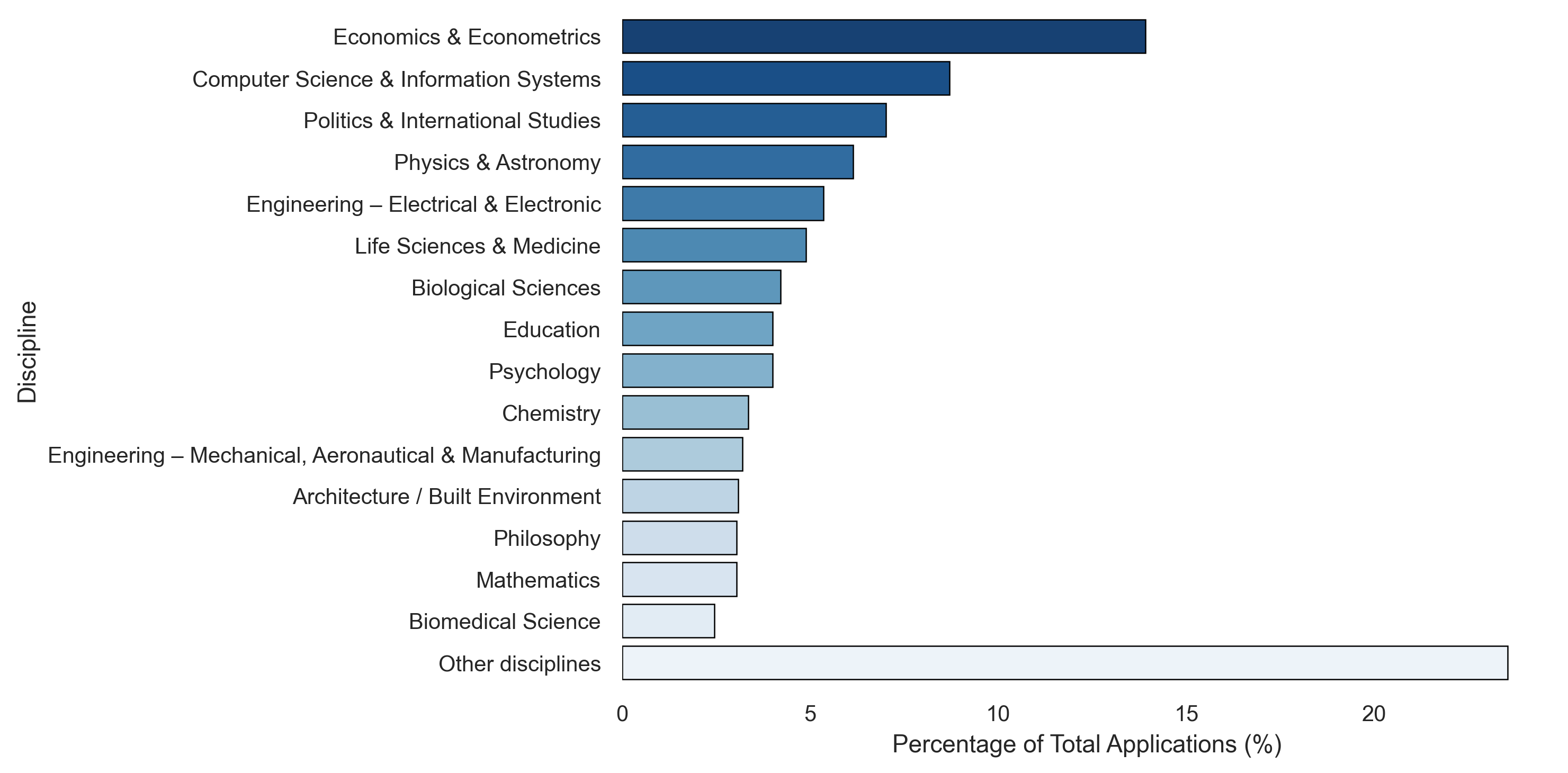}
    \caption{Distribution of applicants across academic disciplines}
    \label{fig:discipline_distribution}
\end{figure}

\begin{figure}[H]
    \centering
    \includegraphics[width=0.5\linewidth]{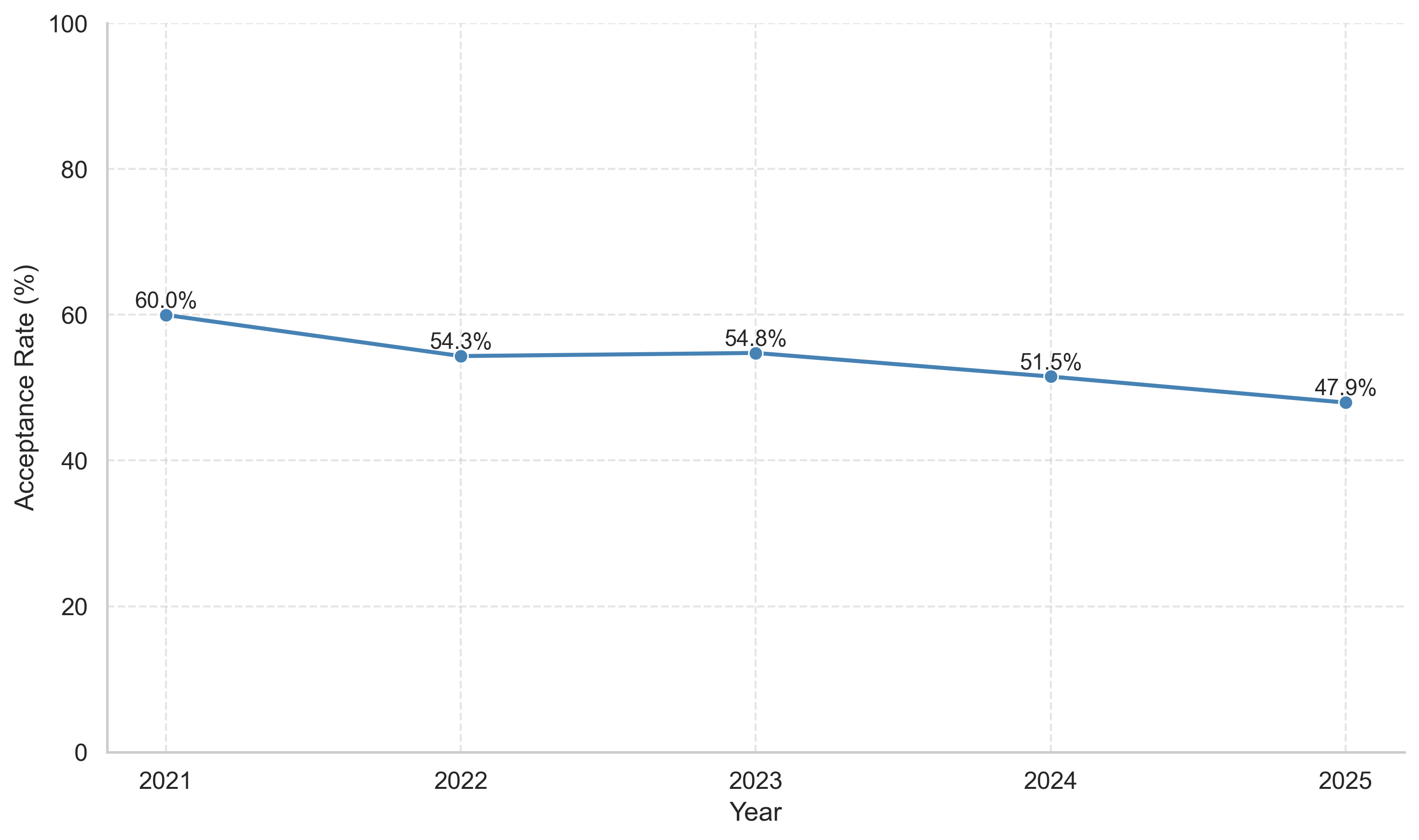}
    \caption{Graduate admission acceptance rate over time}
    \label{fig:acceptance_rate}
\end{figure}

The graduate admission acceptance rate exhibits a gradual decline over the observed period. This trend indicates a modest but consistent increase in competitiveness over time.
\begin{table}[H]
\centering
\arrayrulecolor{blue} 
\renewcommand{\arraystretch}{1.2} 
\begin{tabular}{l r}
\hline
\textbf{Characteristic} & \textbf{Percentage (\%)} \\
\hline
\multicolumn{2}{l}{\textit{University Type}} \\
U.S. universities & 84.53 \\
Non-U.S. universities & 15.47 \\
Public universities & 54.62 \\
Private not-for-profit universities & 44.88 \\
Private for-profit universities & 0.51 \\
\hline
\multicolumn{2}{l}{\textit{Applicant Origin}} \\
International applicants who applied to U.S. universities & 41.42 \\
International applicants who applied to non-U.S. universities & 10.57 \\
American applicants who applied to U.S. universities & 43.12 \\
American applicants who applied to non-U.S. universities & 4.90 \\
American applicants & 48.02 \\
International applicants & 51.98 \\
\hline
\multicolumn{2}{l}{\textit{Acceptance Rate}} \\
Acceptance rate for U.S. universities & 51.73 \\
Acceptance rate for non-U.S. universities & 64.00 \\
\hline
\multicolumn{2}{l}{\textit{Application Type}} \\
Master’s applications & 36.79 \\
PhD applications & 63.21 \\
\hline
\end{tabular}
\caption{Composition of applicants and universities in the dataset}
\label{tab:applicant_university_composition}
\end{table}

\noindent \textit{Percentages indicate the distribution of applicants and universities by type, origin, acceptance rate, and application category.}

\vspace{1em}
\noindent The final processed dataset consists of 13,000 applicants with clean, enriched, coded, and scaled features. Missing values were carefully handled. The observed bias toward U.S. institutions is explicitly acknowledged and should be considered when interpreting results.
\section{Methodology}

\subsection{Problem Formulation and Experimental Setup}

The graduate admission prediction task is formulated as a binary classification problem, aiming to predict whether an applicant will be accepted (1) or rejected (0) for a specific university program degree combination. The model is designed not only to provide label predictions but also to produce calibrated acceptance probabilities, which serve as the quantitative basis for subsequent recommendation strategies.

The dataset split into training (70\%), validation (15\%), and test (15\%) sets using stratified sampling with respect to the target variable to preserve class balance. The resulting set sizes were 9,100 for training, 1,950 for validation, and 1,950 for testing. All stochastic processes, including data splitting, cross-validation, dimensionality reduction, and model initialization, were controlled with a fixed random seed (random\_state = 42) to ensure reproducibility.

Model performance was evaluated using Accuracy, ROC-AUC, Precision, Recall, and F1-score, capturing complementary aspects of predictive correctness, ranking ability, and class wise error behavior. Hyperparameter tuning was performed via 9 fold cross-validation on the training set. The validation set was used solely for model comparison and overfitting control, while final results were reported exclusively on the held out test set.

\subsection{Baseline Models and Selection Rationale}

Linear models offer interpretability but have limited capacity for capturing nonlinear relationships. Tree based models can naturally encode threshold based rules and feature interactions, which are often relevant in admission decisions, such as GPA cutoffs or institutional selectivity. Ensemble methods improve robustness by reducing variance. CatBoost was not included because categorical features had already been pre-encoded, eliminating its primary advantage.
\subsection{Baseline Model Performance Comparison}

Model evaluation employed Accuracy, ROC-AUC, Precision, Recall, and F1-score, all computed on the held out test set. Accuracy measures overall classification correctness, while ROC-AUC evaluates the model's ability to rank accepted and rejected applicants across decision thresholds. Precision reflects the reliability of predicted acceptances, Recall captures the proportion of truly accepted applicants correctly identified, and the F1-score provides a balanced summary of Precision and Recall.

\begin{table}[H]
\centering
\arrayrulecolor{blue} 
\renewcommand{\arraystretch}{1.2} 
\begin{tabular}{l r r r r r}
\hline
\textbf{Model} & \textbf{Accuracy} & \textbf{ROC-AUC} & \textbf{F1-score} & \textbf{Precision} & \textbf{Recall} \\
\hline
Logistic Regression & 0.6944 & 0.7460 & 0.69 & 0.70 & 0.69 \\
Decision Tree & 0.6944 & 0.7460 & 0.69 & 0.70 & 0.69 \\
Random Forest  & 0.7344 & 0.8009 & 0.73 & 0.73 & 0.73 \\
SVM (RBF) & 0.6954 & 0.7365 & 0.69 & 0.69 & 0.69 \\
k-Nearest Neighbors & 0.6774 & 0.7420 & 0.68 & 0.68 & 0.68 \\
XGBoost & 0.7241 & 0.7911 & 0.74 & 0.74 & 0.74 \\
LightGBM & 0.7262 & 0.7927 & 0.73 & 0.73 & 0.73 \\
\hline
\end{tabular}
\caption{Performance comparison of baseline models on the test set across multiple evaluation metrics.}
\label{tab:baseline_performance}
\end{table}

The results indicate that ensemble tree-based models, particularly Random Forest, XGBoost, and LightGBM, consistently outperform simpler linear and distance-based models in terms of accuracy, ROC-AUC, and F1-score. These models effectively capture nonlinear relationships and feature interactions inherent in our dataset. Based on the results in Table~\ref{tab:baseline_performance}, XGBoost was selected as the primary predictive model. While Random Forest achieved slightly higher accuracy and LightGBM demonstrated competitive ROC-AUC, XGBoost provides the best balance between predictive performance, scalability, and interpretability.
\subsection{Test Set Performance and Feature Space Visualization}

\begin{figure}[H]
    \centering
    \includegraphics[width=0.8\linewidth]{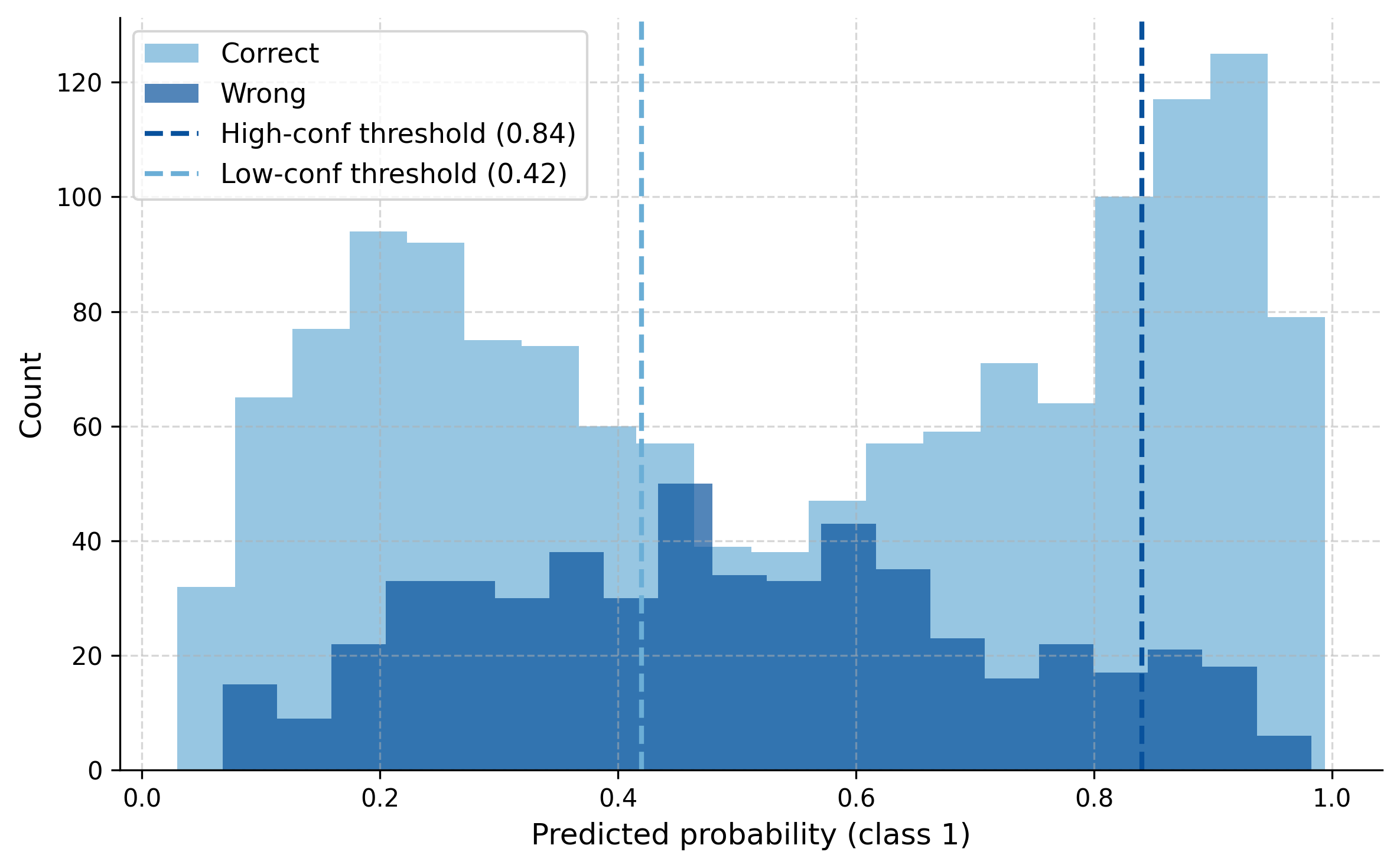}
    \caption{Predicted admission probabilities for all test set samples}
    \label{fig:predicted_probability}
\end{figure}

The XGBoost model was evaluated on the held-out test set of 1,950 samples, producing 1,415 correct predictions and 535 misclassifications. Among the errors, 253 were false positives (predicted as admitted but actually rejected) and 282 were false negatives (predicted as rejected but actually admitted). The predicted probabilities for all test samples are shown in Figure~\ref{fig:predicted_probability}. These probabilities reflect the model's confidence in its predictions, with values closer to 1 indicating high confidence in admission and values closer to 0 indicating high confidence in rejection. Examining the distribution of predicted probabilities helps identify borderline cases where the model is less certain.

To better understand model behavior and class separability, Principal Component Analysis (PCA) was applied. PCA is a linear dimensionality reduction technique that projects the data onto directions (principal components) capturing the maximum variance. By reducing dimensionality, PCA provides a global overview of the structure of the data and the relationships among samples.The test set was projected onto the first two principal components (PC1 and PC2).
\begin{figure}[H]
    \centering
 \includegraphics[width=0.8\linewidth]{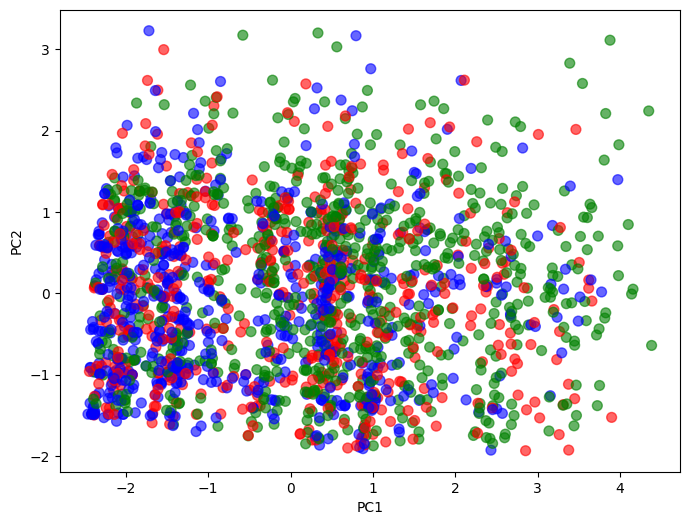}
\caption{PCA projection of test set samples onto the first two principal components.\\
Blue: correctly classified Rejected applicants;\\
Green: correctly classified Accepted applicants;\\
Red: misclassified samples.}

    \label{fig:pca_projection}
\end{figure}

The PCA projection reveals that correctly classified samples cluster in coherent regions where the model is confident, while misclassified samples are dispersed near decision boundaries. Most errors occur for borderline applicants rather than for clearly separable subgroups. Linking PCA with predicted probabilities shows that misclassified samples generally correspond to cases where the model’s confidence is moderate, indicating inherent ambiguity in the data. This suggests that the observed errors and misclassifications are systematic rather than due to sparse data or specific subgroups. Consequently, even removing these borderline samples would not substantially improve model performance. These findings highlight the potential benefit of a hybrid modeling approach, where a residual refinement mechanism could specifically target borderline cases to further enhance predictive accuracy.
\subsection{Hybrid XGBoost + kNN Model}

To improve classification performance beyond what a single global model can achieve, we implemented a hybrid framework combining XGBoost with a local, instance based classifier to refine predictions in challenging cases. Specifically, we define \textit{residuals} as samples where XGBoost either misclassifies the label or predicts with low confidence (borderline probabilities). For these residuals, we considered two alternative local models: k-Nearest Neighbors (kNN) and Support Vector Machine (SVM) with RBF kernel. 

The rationale behind this choice is that residual samples represent borderline or ambiguous applicants, where global patterns captured by XGBoost may fail. kNN can leverage local neighborhood information to adjust predictions based on similarity to nearby applicants, whereas SVM can draw non linear decision boundaries to refine borderline cases. Both approaches aim to improve classification accuracy on the challenging residual subset while preserving the global trends captured by XGBoost.

\begin{table}[H]
\centering
\arrayrulecolor{blue} 
\begin{tabular}{lcccc}
\hline
\textbf{Model} & \textbf{Accuracy (CA)} & \textbf{F1-score} & \textbf{Precision} & \textbf{Recall} \\
\hline
kNN & 0.6774 & 0.69 & 0.71 & 0.67 \\
SVM (RBF) & 0.6779 & 0.68 & 0.70 & 0.70 \\
XGBoost & 0.7242 & 0.74 & 0.74 & 0.74 \\
XGBoost + kNN (Hybrid) & 0.8723 & 0.87 & 0.87 & 0.87 \\
XGBoost + SVM (Hybrid) & 0.8749 & 0.87 & 0.88 & 0.87 \\
\hline
\end{tabular}
\caption{ test set performance metrics}
\arrayrulecolor{black}
\label{tab:overall_metrics}
\end{table}

\bigskip
\noindent
Having selected kNN as the residual model, the hybrid framework can be formally described using the following equations. 
\newpage
Let $y_i$ be the true label of applicant $i$ and $\hat{p}_i$ the XGBoost-predicted probability. The residual set $\mathcal{R}$ is defined as:

\begin{equation}
\mathcal{R} = \{ i \mid (\hat{y}_i^{\text{XG}} \neq y_i) \ \text{or} \ (\hat{p}_i \in [0.4, 0.6]) \},
\end{equation}

where false positives and false negatives are:

\begin{align}
\text{FP} &= \{ i \mid \hat{y}_i^{\text{XG}} = 1 \ \text{and} \ y_i = 0 \}, \\
\text{FN} &= \{ i \mid \hat{y}_i^{\text{XG}} = 0 \ \text{and} \ y_i = 1 \}.
\end{align}

The hybrid prediction for each applicant is then:

\begin{equation}
\hat{y}_i^{\text{Hybrid}} =
\begin{cases} 
\hat{y}_i^{\text{kNN}}, & \text{if } i \in \mathcal{R}, \\
\hat{y}_i^{\text{XG}}, & \text{otherwise},
\end{cases}
\end{equation}

\vspace{1em}
This process is illustrated in Figure~\ref{fig:hybrid_flowchart}:
\vspace{1em}

\begin{figure}[H]
\centering
\begin{tikzpicture}[scale=0.85, node distance=1.5cm, every node/.style={font=\footnotesize}, 
    arrow/.style={->, thick, blue}]
    
\node[draw, rounded corners, minimum width=2.5cm, minimum height=0.8cm] (input) {Applicants' records};
\node[draw, rounded corners, below of=input] (xgb) {XGBoost Prediction};
\node[draw, rounded corners, below of=xgb] (residual) {Residual?};
\node[draw, rounded corners, right of=residual, xshift=4cm] (knn) {kNN Prediction};
\node[draw, rounded corners, below of=residual, yshift=-1.2cm] (final) {Final Hybrid Prediction};

\draw[arrow] (input) -- (xgb);
\draw[arrow] (xgb) -- (residual);
\draw[arrow] (residual.east) -- node[above]{Yes} (knn.west);
\draw[arrow] (knn.south) |- (final.east);
\draw[arrow] (residual.south) -- node[right]{No} (final.north);

\end{tikzpicture}
\caption{Flowchart of the hybrid XGBoost + kNN prediction pipeline}
\label{fig:hybrid_flowchart}
\end{figure}
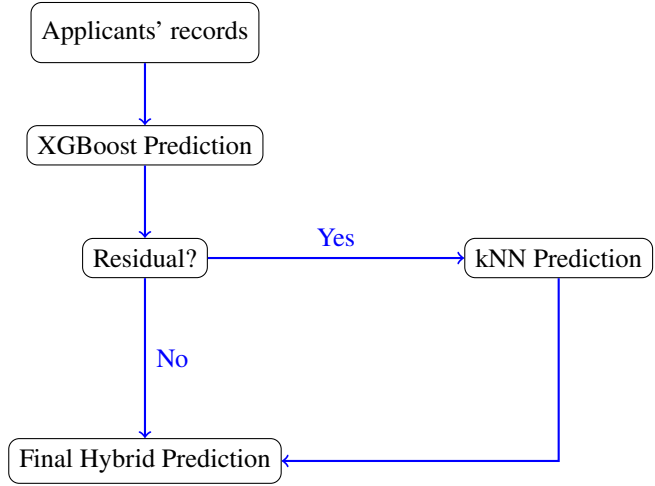

\subsection{Calibration and Final Model Configuration}

\begin{table}[H]
\centering
\arrayrulecolor{blue}
\begin{tabular}{l l l}
\hline
\textbf{Component} & \textbf{Parameter / Metric} & \textbf{Value} \\
\hline
Calibration & Method & Platt scaling (sigmoid) \\
 & Brier score & 0.1864 (Platt), 0.1867 (Isotonic) \\
XGBoost Hyperparameters & colsample\_bytree & 0.8 \\
 & subsample & 0.8 \\
 & reg\_alpha (L1) & 0.1 \\
 & reg\_lambda (L2) & 1 \\
 & n\_estimators & 100 \\
 & learning\_rate & 0.1 \\
 & max\_depth & 8 \\
kNN Hyperparameters & n\_neighbors & 9 \\
 & weights & uniform \\
\hline
\end{tabular}
\caption{Summary of calibration}
\arrayrulecolor{black}
\label{tab:final_config}
\end{table}

Following this configuration, the XGBoost model was calibrated using Platt scaling to produce reliable probability estimates for residual selection.
\subsection{Key Drivers of Admission Decisions Identified by SHAP}

 SHAP quantifies each feature's contribution to the prediction, showing both magnitude and direction. Higher mean $|\text{SHAP}|$ values indicate stronger influence, while positive (negative) values increase (decrease) the likelihood of admission. In our model, the applicant's degree type has the largest positive effect, followed by the applicant's GPA and the program discipline, whereas program level admission rate and program QS ranking reduce the probability of acceptance. These results highlight the key factors driving admission decisions in a transparent, interpretable manner.

\begin{figure}[H]
    \centering
    \includegraphics[width=0.6\linewidth]{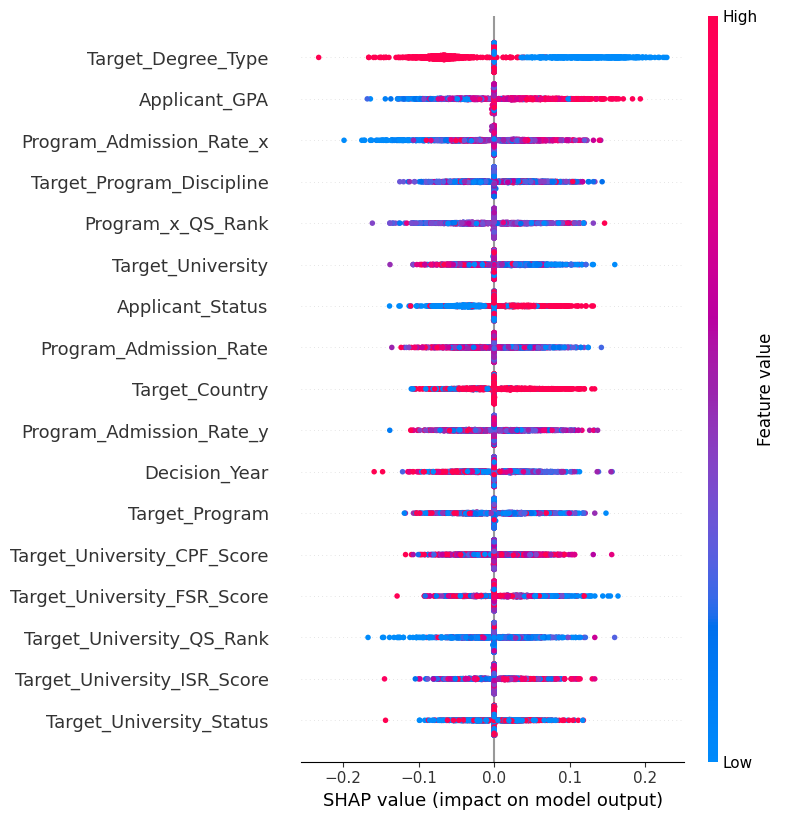}
    \caption{SHAP summary plot showing the impact of each feature on hybrid XGBoost + kNN model predictions. Color indicates feature value from low (blue) to high (red), and the x-axis shows SHAP values (impact on model output).}
    \label{fig:shap_summary}
\end{figure}

\begin{figure}[H]
    \centering
    \includegraphics[width=0.6\linewidth]{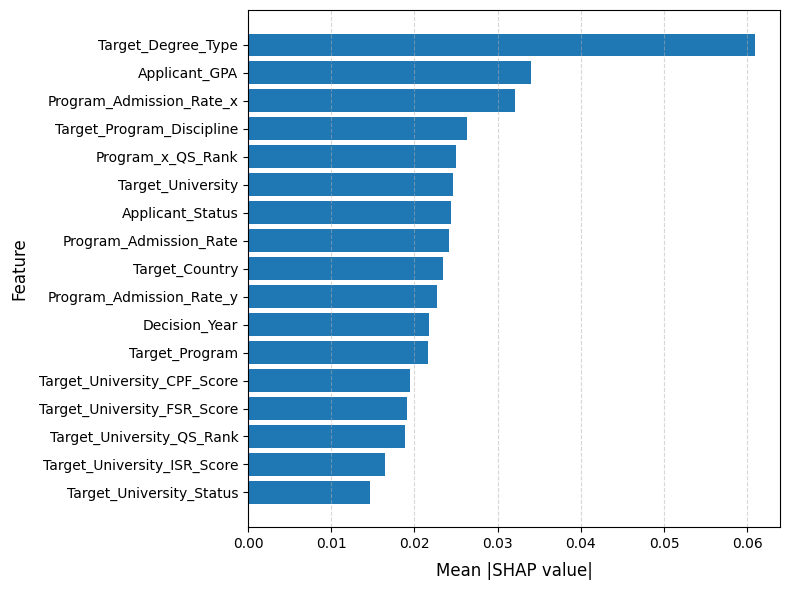}
    \caption{Mean $|\text{SHAP}|$ values for each feature, representing the overall importance of features in driving admission decisions in the hybrid XGBoost + kNN model.}
    \label{fig:shap_importance}
\end{figure}
\subsection{University Recommendation}
University recommendations target applicants who were rejected from their desired institutions. Only candidate universities offering the same program are considered, provided they meet GPA feasibility, affordability, and country preferences. Candidates are ranked primarily by predicted acceptance probability, with GPA proximity, country preference, and budget considerations applied sequentially. The top five universities with the highest predicted probabilities are selected, demonstrating the critical role of institutional choice in admission outcomes.

\subsection{Program Recommendation}
Program recommendations apply a similar approach while keeping the university fixed. Alternative programs within the same discipline are evaluated only if they offer a higher predicted acceptance probability than the original program. Since the university is fixed, country and budget filters are not applied. This method helps applicants targeting competitive institutions identify programs within their targeted descipline with the highest admission likelihood.

\subsection{Hybrid University and Program Recommendation}
The hybrid strategy combines both university and program optimization. Starting from the top universities identified in the university only strategy, alternative programs within each university are assessed. If a program offers a higher predicted probability than the university-level recommendation, the final probability is updated. This approach explicitly maximizes acceptance probability and generates a ranked list of university–program pairs that consistently outperform university only recommendations.

\subsection{Filters and Matching Criteria}
To ensure academic, financial, and geographic feasibility, strict filters are applied. First, applicants are compared to candidate universities based on their QS rank. The maximum allowable GPA difference depends on the QS rank as follows:
\[
\Delta_\text{GPA}^{\max} =
\begin{cases} 
0.10 & \text{if QS} \le 50, \\
0.15 & \text{if QS} \le 150, \\
0.20 & \text{if QS} \le 300, \\
0.25 & \text{if QS} > 300.
\end{cases}
\]

Next, a GPA proximity score is computed to refine candidate ranking. This score ranges from 1 for an exact match to 0 at the maximum allowed difference, and is defined as:
\[
\text{GPA\_Proximity} = 1 - \frac{|\text{Applicant\_GPA} - \text{University\_GPA\_Requirement}|}{\Delta_\text{GPA}^{\max}}
\]
where values are clipped to the range [0,1].

Affordability filters ensure that recommendations are consistent with the applicant's originally targeted university type. Applicants targeting Public universities are recommended only other Public or Private Not-for-Profit (NFP) institutions. Applicants targeting Private NFP universities are recommended either Public or Private NFP alternatives, excluding Private For-Profit (FP) universities. Applicants targeting Private FP universities can receive recommendations from any university type.

Finally, country preference prioritizes universities located in the applicant’s original country. Together, these filters guarantee that all recommended alternatives are academically suitable, financially feasible, and contextually relevant, while maximizing the predicted acceptance probability.

\subsection{Recommendation System Results}
Across the full set of 6,028 rejected applicants, the hybrid recommendation system successfully matched approximately 93\% of applicants under the filters. Considering all rejected applicants, the system increased predicted acceptance probabilities by an average of 65.1\%. Among those matched to at least one better alternative:

\bigskip

\begin{table}[htbp]
\centering
\arrayrulecolor{blue}
\resizebox{\textwidth}{!}{%
\begin{tabular}{lll}
\hline
\textbf{Recommendation Strategy} & \textbf{Average Increase in Acceptance Probability} & \textbf{Applicants with Improvement} \\
\hline
Program-only & 0.142 & 36.57\% \\
University-only & 0.687 & 100\% \\
Hybrid University–Program & 0.704 & 100\% \\
\hline
\end{tabular}%
}
\arrayrulecolor{black}
\caption{Comparison of different recommendation strategies}
\label{tab:recommendation_results}
\end{table}

\section{Discussion and Conclusion}
This study demonstrates that hybrid machine learning models with domain aware feature engineering can effectively predict  competitive graduate admissions and provide actionable recommendations.
Future work includes exploring alternative models like LightGBM and SVM for larger datasets, leveraging more diverse data for scalability, developing real time recommendation interfaces, and extending the logic to richer datasets.

\section*{Acknowledgments}
The author thanks GradCafe, OpenAlex, WikiData, and QS World University Rankings for providing open source data, and MohammadReza Afshar for collecting the raw dataset through web scraping.

\bibliographystyle{IEEEtran}
\bibliography{References}

\end{document}